\documentclass[aps,prd,preprint,superscriptaddress,showpacs,floatfix, preprintnumbers,nofootinbib,
a4paper]{revtex4-1}
\usepackage{color}
\usepackage{graphicx}
\usepackage{textcomp}
\usepackage{amsmath,amssymb}
\usepackage{enumerate}
\usepackage{hyperref}
\hypersetup{
  colorlinks=true,
  citecolor=blue,
  linkcolor=blue,
  urlcolor=blue}
\newcommand{\mathsym}[1]{{}}

\newcommand{\ba}{\begin{array}}
\newcommand{\ea}{\end{array}}
\newcommand{\be}{\begin{equation}}
\newcommand{\ee}{\end{equation}}
\newcommand{\beqa}{\begin{eqnarray}}
\newcommand{\eeqa}{\end{eqnarray}}
\def\321{$SU(3)\times SU(2)\times U(1)$}

\def\vev#1{\left\langle #1\right\rangle}

\newcommand{\Dms}{\Delta m^2_{\rm sol}}
\newcommand{\Dma}{\Delta m^2_{\rm atm}}

\begin{document}
\vspace*{1cm}
\title{Neutrino masses and mixing in  $A_5$ with flavour antisymmetry}
\author{Anjan S. Joshipura}
\email{anjan@prl.res.in}
\affiliation{Physical Research Laboratory, Navarangpura, Ahmedabad 380 009, India.}
\author{Newton Nath}
\email{newton@prl.res.in}
\affiliation{Physical Research Laboratory, Navarangpura, Ahmedabad 380 009, India.}
\affiliation{Indian Institute of Technology, Gandhinagar, Ahmedabad--382424, India.}
 \vspace*{1cm}

\begin{abstract}
We discuss consequences of assuming  that the (Majorana)
neutrino mass
matrix $M_\nu$ and the charged lepton mass matrix $M_l$ satisfy,
$S_\nu^T M_\nu S_\nu=-M_\nu~,T_l^\dagger M_lM_l^\dagger T_l=M_lM_l^\dagger$ 
with respect to some discrete groups $S_\nu,T_l$ contained in $A_5$.
These assumptions lead to a neutrino mass spectrum with two degenerate and one massless neutrino and also 
constrain mixing among them. We derive possible  mixing patterns following
from the choices
$S_\nu=Z_2,~ Z_2\times Z_2$ and $T_l=Z_2, ~ Z_2\times Z_2, ~Z_3, ~Z_5$ as subgroups of $A_5$. One predicts  the maximal atmospheric neutrino mixing angle $\theta_{23}$ and $\mu$-$\tau$ reflection symmetry in large number of cases but it is also possible to obtain  non-maximal values for $\theta_{23}$. Only the third column of the neutrino mixing matrix can be obtained at the leading order due to degeneracy in masses of two of the neutrinos. We take up a specific example within
$A_5$ group and identify Higgs vacuum expectations values which realize the above assumptions. Non-leading terms present in this example are shown to lead to splitting among degenerate pairs and a consistent description of both neutrino masses and mixing angles.

\vspace*{0.2cm}
\end{abstract}
\maketitle
\section{Introduction}
Two decades of neutrino oscillation experiments have determined five
of the key parameters describing oscillations of three neutrinos. These are
three mixing angles $\theta_{ij}~, (i,j=1,2,3; i<j)$ and two (mass)$^2$
differences $\Delta_\odot$ and $\Delta_A$ controlling the oscillations of
the solar and the  atmospheric neutrinos respectively. Overall  neutrino
mass scale and three CP violating phases still remain to be determined.
There already exists hint that the CP phase $\delta$ may be nearly maximal. 

Theoretical frameworks describing neutrino masses and mixing angles try to
understand the values of the observed parameters and aim to predict the
unknown ones. Flavour symmetries provide concrete framework to do this. 
A systematic approach based on flavour symmetries has evolved in last
several
years, see reviews
\cite{Altarelli:2010gt,Altarelli:2012ss,Smirnov:2011jv,Ishimori:2010au,
King:2013eh} and references therein.  This is based on the observation that
patterns of neutrino masses and
mixing is intimately linked to the residual symmetries of the neutrino and
the charged lepton mass matrices \cite{Lam:2007qc, Lam:2008rs,Lam:2008sh}.
These residual symmetries of mass
matrices can be related to the full symmetry $G_f$ of the underlying theory
by assuming that former symmetries are contained in $G_f$. This provides a
direct link between the group theoretical structure of $G_f$ and the
observed mixing angles. This approach has been used to predict various
mixing patterns consistent with observations in large number of cases with
many different  discrete symmetry groups  $G_f$
\cite{Altarelli:2010gt,Altarelli:2012ss,Smirnov:2011jv,Ishimori:2010au,
King:2013eh,Fonseca:2014lfa}. 

The above approach is also generalized to link both the mass and the  mixing
patterns of neutrinos to some underlying symmetries. Three possible neutrino
mass patterns provide a good zeroeth order approximation to the observed
neutrino mass spectrum, fully degenerate spectrum, quasi degenerate spectrum
with two degenerate neutrinos and a spectrum with two massive and one
massless neutrinos. A systematic procedure is evolved to relate these
patterns to underlying discrete symmetries. A general analysis is presented
for  three classes of groups, the discrete von-Dyck groups  in case of the
degenerate and quasi degenerate spectrum \cite{Hernandez:2013vya}, all
possible discrete subgroups 
of $SU(3)$ having 3 dimensional irreducible representation in case of the quasi degenerate neutrinos
\cite{Joshipura:2014qaa}
and a large class of 
discrete subgroups of $U(3)$ in case of one
massless neutrino \cite{Joshipura:2013pga,Joshipura:2014pqa,King:2016pgv} .

The basic assumption in the above approaches is that the underlying theory
is invariant under some discrete group $G_f$  but the Higgs  vacuum expectation value (vev)
determining  neutrino
mass matrix $M_\nu$  and the Hermitian combination of the charged lepton
mass matrix $M_lM_l^\dagger$ remain invariant   under smaller subgroups 
$G_\nu$ and $G_l$ of $G_f$. The structure of these groups and their
embedding in $G_f$
is sufficient for the  determination of mixing patterns
 without the knowledge of the detailed 
dynamics.  A different dynamical possibility was studied in
\cite{Joshipura:2015zla}. Here it is
assumed that  the Higgs vacuum expectation values breaking flavour group $G_f$  lead to a neutrino mass matrix
which displays antisymmetry. Specifically, $M_\nu$ satisfies  
\be \label{flavas}
S_\nu^T M_\nu S_\nu=-M_\nu~\ee
for some subgroups $S_\nu$ of $G_f$. This assumption was shown \cite{Joshipura:2015zla} to constrain not just the mixing angles but also
 the neutrino mass spectrum which 
could be determined purely from the group theoretical arguments.
Detailed mixing and mass patterns allowed within the discrete subgroups
$\Delta(3N^2)$ and $\Delta(6N^2)$ and a  specific dynamical realization of
the basic idea in case of the group $A_4\equiv \Delta(3. 2^2)$ was discussed
in \cite{Joshipura:2015zla}. Also it was shown in a specific example that the antisymmetry of the mass matrix can arise from the minimization of some suitable potential.  Here we pursue this idea further and apply it
to the symmetry group $A_5$. We discuss mass patterns and all the mixing
patterns possible within $A_5$ using the idea of flavour antisymmetry of
neutrino mass matrix. $A_5$ has been used in the past
\cite{Everett:2008et,Feruglio:2011qq,Ding:2011cm,
Ding:2011cm,Varzielas:2013hga} to
predict the
neutrino mixing patterns assuming flavour symmetry. The
mixing patterns predicted here are quite different compared to these cases.

Detailed analysis of $A_5$ also becomes interesting from a related point of
view. It was shown \cite{Joshipura:2015dsa} that all the discrete subgroups
of $O(3)$
can lead to universal prediction $\theta_{23}=\frac{\pi}{4}$ and $|\delta|=\frac{\pi}{2}$ when $G_\nu$  is chosen as $Z_2\times Z_2$ or $Z_m$ and
$G_l$ is chosen as $Z_n$, $m,n\geq 3$. As we will see, the same predictions also follow  when neutrino mass matrix possesses residual antisymmetry instead of symmetry.

We review in the next section some of the properties of the group $A_5$ relevant for our study.
We introduce the idea of flavour antisymmetry in section 3 and discuss its consequences. Section 4 is devoted to a detailed discussion of various mixing patterns possible within the group $A_5$ under the assumption of the flavour antisymmetry. Section 5 discusses explicit realization of the ideas discussed in the previous section. The last section summarizes the findings.

\section{$A_5$ and its Abelian subgroups}
Group theory of $A_5$ is discussed in several papers
\cite{Everett:2008et,Feruglio:2011qq,Ding:2011cm,
Ding:2011cm,Grimus:2011fk,Varzielas:2013hga}. We summarize here the features
which we require for subsequent analysis.
The $A_5$ group has sixty elements and five conjugacy classes. The group  can be  represented in terms of three generators 
$E,f_1,H$,
\be \label{def}
H=1/2\left(\ba{ccc}
-1&\mu_ -&\mu_+\\
\mu_-&\mu_+&-1\\
\mu_+&-1&\mu_-\\ \ea \right) ~~~;~~~E=\left(\ba{ccc}
0&1&0\\
0&0&1\\
1&0&0\\ \ea \right)~~~;~~~f_1=\left(\ba{ccc}
1&0&0\\
0&-1&0\\
0&0&-1\\
 \ea \right)~,\ee
with $\mu_\pm=1/2(-1\pm \sqrt{5})$ which provide a faithful 3 dimensional irreducible 
representation. The above equation defines the basis of the representation
labeled as $3_1$ and we will refer to this basis as symmetry basis. Multiple
products of these generate all the sixty elements of $A_5$. It is convenient
for our purpose to discuss these elements in terms of the $Z_n$ subgroups
they form.
We list them and their required properties below.
 
\begin{itemize}
\item $\mathcal{ Z}_2$: 15 $Z_2$ subgroups of $A_5$ are generated by the  elements:
\be \label{o2}
O_2\equiv (f_a, H, f_a Hf_a, EHE^{-1},E^{-1}HE, Ef_aHf_aE^{-1},E^{-1}f_aHf_a E)~,\ee
where $a=1,2,3$, $f_2=E^2f_1E,f_3=E^2f_2E$ and $f_1$ is given by eq.(\ref{def}).
One also needs the matrices which diagonalize the elements in $O_2$ when $Z_2$ is used as a residual symmetry.
These get determined by a matrix $V_H$ which diagonalizes $H$.  Let $V_H$
be such matrix then
\be \label{uhd}
V_H^\dagger H V_H={\rm diag.}(1,-1,-1) ~.\ee
Explicitly,
\be \label{vh}
V_H=\left(\ba{ccc}
\frac{1}{2}&-\frac{\sqrt{3}}{2}&0\\
\frac{\mu_-}{2}&\frac{\mu_-}{2\sqrt{3}}&\frac{\mu_+}{\sqrt{3}}\\
\frac{\mu_+}{2}&\frac{\mu_+}{2\sqrt{3}}&-\frac{\mu_-}{\sqrt{3}}\\
\ea
\right)
\ee
The above $V_H$ is arbitrary upto a unitary rotation in the 23 plane. We shall use the above explicit form
for the subsequent analysis.
We can express all the elements of $A_5$ in the form
$QPQ^{-1}$.
This simplifies their  diagonalization  since  $U_{QPQ^{-1}}=QU_P$ where,
$U_g$ diagonalizes  the element $g$. Using this, the 
matrices diagonalizing all the 15 elements  in $O_2$ can be expressed in
terms of $V_H$ and are given by the following set
\be\label{u2}
\mathcal{U}_2\equiv (I,V_H,f_aV_H,EV_H,E^{-1}V_H,Ef_aV_H,E^{-1}f_aV_H)~.\ee
Respective entries of this set correspond to matrices which diagonalize the
corresponding elements of $O_2$. 
\item $\mathcal{Z}_2\times \mathcal{Z}_2:$ Not all the fifteen elements in $O_2$ commute among themselves.
But one can find five sets of three commuting elements among $O_2$. These
three along with identity form a $Z_2\times Z_2$ subgroup of $A_5$. These
subgroups are listed in Table \ref{z2xz2}.
 \begin{table}[!ht]
 \begin{center}
\begin{math}
\begin{tabular}{|c|c|c|c|}
 \hline
 \hline
 $S_1$&$S_2$&$S_3$&$U_c$\\
 \hline
 $f_1$&$f_3$&$f_2$&$I$\\
$H$&$E^{-1}f_2Hf_2E$&$Ef_3Hf_3E^{-1}$&$V_HR_\mu$\\
$f_1Hf_1$&$E^{-1}f_1Hf_1E$&$Ef_1Hf_1E^{-1}$&$f_1V_HR_\mu$\\
\hline
$f_2Hf_2$&$E^{-1}f_3Hf_3E$&$EHE^{-1}$&$f_2V_HR_\mu$\\
\hline
$f_3Hf_3$&$E^{-1}HE$&$Ef_2Hf_2E^{-1}$&$f_3V_HR_\mu$\\
\hline
\hline
\end{tabular}
\end{math}
\end{center}
\caption{Elements of the  five $Z_2\times Z_2$ subgroups of $A_5$ along with their combined
diagonalizing matrices $U_c$ defined in the text. $S_1,S_2,S_3$ together with
identity form a $Z_2\times Z_2$ subgroup of $A_5$.}
\label{z2xz2}
\end{table}
Since $S_1$ and $S_2$ in the table commute, they can be simultaneously diagonalized by a matrix $U_c$. We shall define $U_c$ as
\beqa
\label{uc}
U_c^\dagger S_1U_c=f_1={\rm diag.}(1,-1,-1)~,\nonumber\\
U_c^\dagger S_2U_c=f_3={\rm diag.}(-1,-1,1)~.\eeqa
The same matrix $U_c$ also puts $S_3=S_1S_2$ into a diagonal form $f_2$. As before, the matrix $U_c$  can also be expressed in terms of $V_H$ diagonalizing $H$ and a real rotation $R_\mu$ in the 23 plane  
\be
R_\mu=\left(\ba{ccc}
1&0&0\\
0&-\sin \theta_ \mu&\cos\theta_\mu\\
0&\cos \theta_\mu&\sin \theta_\mu\\
\ea
\right)~,\ee
where
$$\tan \theta_\mu=\mu_- -1~.$$
$U_c$ for all five subgroups are given in Table \ref{z2xz2}. 
\item {$\mathcal{Z}_3$  subgroups:} The 20 elements  generating $Z_3$
subgroups of $A_5$ are given by the set
\be \label{o3}
O_3=(E^m,f_aE^mf_a,A^m,EA^mE^{-1},E^{-1}A^mE,AE^mA^{-1},Af_{2,3}E^mf_{2,3}A^
{-1})~.\ee
$m=1,2$,$a=1,2,3$ and the matrix $A\equiv Hf_1$. The matrices diagonalizing
these elements can be expressed in terms of the matrices $U_\omega$ and
$U_A$ which diagonalize $E$ and $A$ respectively:
\be \label{u3}
\mathcal{U}_3=(U_\omega,f_aU_\omega,U_A,EU_A,E^{-1}U_A,AU_\omega,Af_{2,3}
U_\omega)~.\ee
$U_\omega$ is given by 
\be \label{uw}
U_\omega=\frac{1}{\sqrt{3}}\left(
\ba{ccc}
1&1&1\\
1&\omega&\omega^2\\
1&\omega^2&\omega\\  \ea \right)~,\ee
$\omega=e^{\frac{2 \pi i}{3}}$ and $U_A$ can be found in the Appendix of the
reference \cite{Joshipura:2015dsa}.
\item {$\mathcal{Z}_5$  subgroups:}
There are 24 different $Z_5$ subgroups within $A_5$. Their generating
elements can be expressed in terms of $T\equiv f_1EH$, $E$ and $f_{1,2,3}$ as follows:
\be \label{o5}
O_5=(T^p,f_2T^Pf_2,ET^pE^{-1},E^{-1}T^pE,Ef_2T^pf_2E^{-1},E^{-1}f_2T^pf_2E
)~,\ee
where $p=1,2,3,4$. This set is diagonalized by
\be \label{u5}
\mathcal{U}_5=(U_T,f_2U_T,EU_T,E^{-1}U_T,Ef_2U_T,E^{-1}f_2U_T)~,\ee
where $U_T$ is a matrix diagonalizing $T$. Its explicit form is given in
the Appendix of \cite{Joshipura:2015dsa}.
\end{itemize}

The elements in the sets $O_{2,3,5}$ along with the identity constitute all the sixty elements of $A_5$.
We note that all the matrices diagonalizing set $O_3$ and $O_5$ possess the following
general form as explicitly shown in \cite{Joshipura:2015dsa}.
\be \label{ulz3z5}
 U=\left(
\ba{ccc}
x_1&z_1&z_1^*\\
x_2&z_2&z_2^*\\
x_3&z_3&z_3^*\\
\ea \right)~,\ee
where $x_1,x_2,x_3$ are real. We shall use this form to derive properties of the mixing matrix in the following.
\section{Flavour antisymmetry and neutrino mass textures} 
We first briefly review  the implications of the assumption of the flavour
antisymmetry \cite{Joshipura:2015zla} represented by eq.(\ref{flavas}) where $S_\nu$ is assumed to be
any $3\times 3$ matrix belonging to a discrete subgroup of $SU(3)$.
The very assumption of flavour antisymmetry implies that (at least) one of
the neutrinos remains massless. This simply follows by taking the
determinant of 
eq.(\ref{flavas}) and noting that $Det(S_\nu) = 1$. Other implications of eq.(\ref{flavas})  become clear
in a basis with diagonal $S_\nu$. Let  $S_\nu$ be diagonalized by a unitary
matrix $V_{S_\nu}$ as :
\be 
\label{snudiag}
V_{S_\nu}^\dagger S_\nu V_{S_\nu}=D_S\equiv {\rm diag.}
(\lambda_1,\lambda_2,\lambda_3)~,\ee
with $\lambda_1\lambda_2\lambda_3=1.$
Unitarity of $S_\nu$ implies that $\lambda_{1,2,3 }$ are some 
roots of unity. It was argued \cite{Joshipura:2015zla} that only two
possible forms of $D_S$ can lead to a neutrino mass matrix with two massive
neutrinos. These are given by
\beqa \label{dsforms}
D_{1S}&=&{\rm diag.} (\lambda,-\lambda^*,-1)~,\nonumber \\
D_{2S}&=&{\rm diag.}(\pm i,\mp i,1)~,\eeqa
and their permutations. $\lambda^{2p}=1$ for some integer $p$\footnote{Note that eq.(\ref{flavas}) requires
that $S_\nu^{2 p}=1$ if $M_\nu$ is not identically zero and $S_\nu$ has finite order. This translates to $\lambda^{2 p}=1$.}. The group
generated by  the residual symmetry $S_\nu$ having diagonal form $D_{1S}$
($D_{2S}$) is $Z_{2p}$($Z_4$). 
Define 
\be \label {mtilde}
\tilde{M}_\nu=V_{S_\nu}^TM_\nu V_{S_\nu}~.\ee
Then the allowed textures of $\tilde{M}_\nu$ get determined by the allowed
forms of $D_S$. There exists only four allowed textures for $\tilde{M}_\nu$
which correspond to one massless and a degenerate or non-degenerate pair of
neutrinos. If $\lambda=1$ then the relevant texture is given by
:
\be \label{degtext}
\ba{ccc}
\tilde{M}_\nu=m_0\left(\ba{ccc} 0&c_\nu&s_\nu e^{i\beta_\nu}\\
c_\nu&0&0\\
s_\nu e^{i\beta_\nu}&0&0\\
\ea\right)\\
\ea
~,\ee
where $c_\nu=\cos\theta_\nu,s_\nu=\sin\theta_\nu$. $\tilde{M_\nu}$ describes
a
massless and
a degenerate pair of neutrinos.
 Other three  textures are possible for other values of $\lambda$ but as we
shall see only the case given in eq.(\ref{degtext}) can get realized in
$A_5.$ 
\subsection{The allowed residual symmetries in $A_5$}
We now discuss  possible residual symmetries of the leptonic mass matrices
within  $A_5$ and the resulting mixing patterns. The choices of residual
antisymmetry of $M_\nu$ within $A_5$ are restricted. These 
can be  obtained simply from the characters $\chi$ of all the sixty
elements. $\chi$ is real for all the elements. In this case, the
eigenvalues of any element are given by
\be
\label{ev}
(1,\frac{1}{2}\left( \chi-1+ \sqrt{(\chi-1)^2-4}\right),\frac{1}{2}\left( \chi-1- \sqrt{(\chi-1)^2-4}\right))~.\ee
These eigenvalues must have the form displayed in one of the two equations
given in (\ref{dsforms}) in order for an element with character $\chi$ to
be able to be a viable antisymmetry operator. Elements belonging to $Z_3$
and $Z_5$ subgroups have $\chi=0$ and $(-\mu_+,-\mu_-)$. Their eigenvalues
following from above do not have  these forms. Thus only viable
choice for the antisymmetry operator $S_\nu$ can be any element in the set
$O_2$ having character -1 and eigenvalues $(1,-1,-1)$. We shall require that at least
one of the symmetries of $M_\nu$ acts according to eq.(\ref{flavas}). We
will thus consider
two possible choices of the residual
neutrino symmetries  (1) $S_\nu=Z_2$ as antisymmetry and (2)
$S_\nu=Z_2\times Z_2$ where one of the $Z_2$ transforms $M_\nu$ into it's
negative and the other leaves it invariant. In contrast, the eigenvalues 
of the  residual symmetry of $M_lM_l^\dagger$ is not restricted and we
can take
any of the $Z_n$ of $A_5$ as the residual symmetry $T_l$. We shall consider the following choices for $T_l$
(a) $(Z_3,Z_5)$ groups generated by $(O_3,O_5$) (b) five  $Z_2\times Z_2$
subgroups  or (c) elements of the $Z_2$ subgroups
contained in $O_2$. Possible choices of $S_\nu$ and $T_l$ 
determine the leptonic mixing matrix.

Elements in $O_2$ when used as
antisymmetry operator lead to a unique form for the neutrino mass matrix
$\tilde{M}_\nu$  given in eq.(\ref{degtext}). This texture describes a pair
of degenerate and one massless neutrino. Residual antisymmetry in this case
is $Z_2$.
The neutrino mass matrix in eq.(\ref{degtext}) can be   diagonalized by a  matrix  $V_\nu$:
\be \label{dm}
V_\nu^T \tilde{M}_\nu V_\nu={\rm diag.}(m_0,m_0,0) \ee
where,
\be \label{unu1}
V_\nu=\left( \ba{ccc}
\frac{1}{\sqrt{2}}&-\frac{i}{\sqrt{2}}&0\\
\frac{c_\nu}{\sqrt{2}}&\frac{i c_\nu}{\sqrt{2}}&-s_\nu\\
\frac{s_\nu}{\sqrt{2}}e^{-i\beta_\nu}&\frac{i
s_\nu}{\sqrt{2}}e^{-i\beta_\nu}&c_\nu e^{-i\beta_\nu}\\
\ea\right)
\left( 
\ba{ccc}
\cos\psi&-\sin\psi&0\\
\sin\psi&\cos\psi&0\\
0&0&1\\
\ea \right)
~.\ee
The arbitrary rotation by an angle $\psi$ originates due to degeneracy in
masses. It follows from eqs.(\ref{mtilde},\ref{dm})  that the
matrix $M_\nu$ is diagonalized by the product $V_{S_\nu}V_\nu$.  Thus the
neutrino mixing matrix with the residual antisymmetry $Z_2$ in the symmetry
basis  is given by
\be\label{unuz2}
U_\nu^I=V_{S_\nu}V_\nu ~.\ee
Note that the $U_\nu^I$ gets determined by the structure of $S_\nu$ and
essentially  two unknown 
angles $\theta_\nu$ and $\beta_\nu$.  The unknowns can be 
fixed if the residual symmetry is chosen as $Z_2\times Z_2$. Consider the
$Z_2\times Z_2$ groups generated by  $S_{1\nu}=S_1$ and $S_{2\nu}=S_2$ where
$S_1,S_2$ are as in Table \ref{z2xz2}. They satisfy
\be \label{s1s2}
S_1^TM_\nu S_1=-M_\nu~~~~~~;~~~~~S_2^TM_\nu S_2=M_\nu~.\ee

As discussed in the previous section, both $S_1$ and $S_2$ are diagonalized
by $U_c$ as given in Table \ref{z2xz2}. The structure of the neutrino mass
matrix in
this case becomes transparent in the basis with diagonal  $S_1,S_2$
Let 
\be \label{mnuprime}
M_\nu^\prime =U_c^T M_\nu U_c ~.\ee
Eq.(\ref{s1s2}) reduces in the prime basis to
\be \label{s1s2prime}
f_1^TM_\nu^\prime f_1=-M_\nu^\prime~~~~~,~~~~~
f_3^TM_\nu^\prime f_3=M_\nu^\prime~.\ee
The first of this equation implies the form (\ref{degtext}) for
$M_\nu^\prime$. The second imposed on this then leads to
the restriction $s_\nu=0,c_\nu=1$. The final $M_\nu^\prime$ is determined by an overall scale $m_0$ and is diagonalized by
$U_{12}\equiv R_{12}(\frac{\pi}{4}) {\rm diag.}(1, i,1)$. It follows from this and eq.(\ref{mnuprime}) that $M_\nu$ is diagonalized by
\be \label{unuz2z2}
U_\nu^{II}=U_cU_{12}=U_c R_{12}(\frac{\pi}{4}) {\rm diag.}(1, i,1)~.\ee

The matrix $U_l$ diagonalizing  $M_lM_l^\dagger$
also gets determined by its symmetry. Assume that
\be
\label{mlsym}
T_l^\dagger M_lM_l^\dagger T_l=M_lM_l^\dagger ~. \ee
This implies that $T_l$ commutes with $M_lM_l^\dagger$. Hence the matrix
$U_{T_l}$ diagonalizing $T_l$ can be taken to be the matrix which
diagonalizes $M_lM_l^\dagger$ also. Three possible choices of $T_l$ referred
as (a),(b),(c) above lead to specific forms of $U_l$:
\beqa
\label{ulabc}
U_l^a&=& U_{3,5}~,\nonumber\\
U_l^b&=&U_c~,\nonumber\\
U_l^c&=&U_2U_{23}~.\eeqa
Here, $U_{3,5}$ are  given by any matrix in the set, $\mathcal{
U}_3$ , eq.(\ref{u3}) and $\mathcal{ U}_5$ , eq.(\ref{u5}) when $T_l$
belongs to
$O_3$ or $O_5$ respectively. $U_c$ is  given in Table \ref{z2xz2} for
$T_l$ belonging to $Z_2\times Z_2$. There is some arbitrariness in the
choice of
$U_l$ when $T_l$ is chosen as any of the element $O_2$ forming a $Z_2$.
These elements have eigenvalues $(1,-1,-1)$ and matrix diagonalizing $T_l$
is arbitrary up to a unitary rotation in the 23 plane. This rotation can be
taken without the loss of generality to
$U_{23}\equiv {\rm diag.}(1,1, e^{i\beta_l}) R_{23}(\theta_l)$. Various
combinations of $U_l^{a,b,c}$ and
$U_\nu^{I,II}$ give all  possible $U\equiv U_l^\dagger U_\nu$ in $A_5$. 
\section{Mixing Patterns in $A_5$}
As discussed, all possible  structure of the PMNS matrix $U$ in $A_5$
with flavour antisymmetry are given by
\be \label{upmns}
U\sim U_l^{\dagger a,b,c} U_\nu^{I,II}~.\ee
Not all of these give viable mixing pattern for neutrinos as we will show.
Before discussing individual choices, we first derive a fairly  general
property of the mixing matrix  with flavour antisymmetry. If  (a) the 
neutrino mass matrix shows flavour antisymmetry eq.(\ref{flavas}), with 
$S_\nu^2=1$ and a real
mixing matrix $V_\nu$ or (b) if  it  has residual symmetry structure
$Z_2\times Z_2$ as in eq.(\ref{s1s2}) and if the charged lepton matrix
$M_lM_l^\dagger$ is invariant under a residual symmetry $Z_3$ or $Z_5$
within $A_5$ then the mixing matrix can be chosen to have the property
\be \label{genmt}
|U_{\mu i}|=|U_{\tau i}|~,~(i=1,2,3)~.\ee
This property known as the $\mu-\tau$ reflection symmetry
\cite{Harrison:2002et} or
generalized $\mu$-$\tau$ symmetry  was derived \cite{Grimus:2003yn}  using
a generalized definition of CP.  The same result was derived from more
general assumptions in case of the non-degenerate neutrinos
\cite{Joshipura:2015dsa,He:2015xha}
as well as for a pair of degenerate neutrinos
\cite{Hernandez:2013vya,Joshipura:2015dsa}. The
basic assumption in these cases was the existence of a real residual
symmetry. The same result also follows when the
symmetry is replaced by antisymmetry as we discuss below.

The equality $|U_{\mu 3}|=|U_{\tau 3}|$ implies the maximal atmospheric
mixing
angle. The equality $|U_{\mu 2}|=|U_{\tau 2}|$
then leads to the maximal CP phase $|\delta|=\frac{\pi}{2}$ if neutrinos are
non-degenerate and $s_{13}\not=0$. For the
degenerate solar
pair, the first two columns of $U$ depend on an unknown mixing angle $\psi$
as given in eq.(\ref{unu1}). But by considering $\psi$
invariant combination of the observables, it was argued
\cite{Hernandez:2013vya} that  one instead
gets $|\delta -\kappa|=\frac{\pi}{2}$ where $\kappa$ is the Majorana phase.

The proof of eq.(\ref{genmt})   is straightforward and follows the proof
given  in \cite{Joshipura:2015dsa} in case of the flavour symmetry. Assume
that neutrino
mass matrix ${\tilde M}_\nu$ has the structure (\ref{degtext}) with
$\beta_\nu=0$. Then it is diagonalized by $U_\nu^I=U_2V_\nu$. Here
$U_2$ belonging  to the set $\mathcal {U}_2$ 
is real. For $\beta_\nu=0$, one therefore gets $U_\nu^I=O_\nu P$, with  $O_\nu$ 
being a real orthogonal matrix and $P=  {\rm diag.}(1,i,1)$. 
A similar 
structure of $U_\nu$ also holds in
case II with $Z_2\times Z_2$ symmetry since in this case, the
neutrino mixing matrix $U_\nu$ is given by $U_\nu^{II}$,
eq.(\ref{unuz2z2}) which also can be written as an orthogonal matrix times
a phase matrix because of the reality of $U_c$. The charged lepton mixing
matrix on the other hand has a general structure
specified by eq.(\ref{ulz3z5}) when the residual charge lepton symmetry is
either $Z_3$ or $Z_5$.  It is then easy to see that $U_l$ as in
eq.(\ref{ulz3z5}) and $U_\nu$ as $O_\nu$ times a diagonal phase matrix leads
to eq.(\ref{genmt}). This result does not follow when the residual symmetry
of the charged leptons is $Z_2$ or $Z_2\times Z_2$ since in this case $U_l$
does not have the form given in eq.(\ref{ulz3z5}).
 
Let us now discuss individual choices of residual symmetries and their
viability or otherwise. We will work out various mixing patterns for various
choices and 
confront them with the results of the global fits as
given in \cite{Forero:2014bxa,Gonzalez-Garcia:2014bfa,Capozzi:2013csa}.
For definiteness, we shall use the results given in \cite{Capozzi:2013csa}.
The structures of $U^{I,II}$ appearing in
eq.(\ref{upmns})  are determined only up to a rotation in the 12 plane and
the solar angle remains undetermined at the leading order. The third column of $U$ is however
independent of the unknown angle and can be predicted group theoretically at
the zeroeth order. We shall thus concentrate on the prediction of
$\theta_{13}$ and $\theta_{23}$ determined by the third column of $|U|$.
Also, the ordering of eigenvalues of $T_l$ cannot be determined group
theoretically. Change in this ordering permutes the rows of $U$. Thus any
of the entries of the third column $|U_{i3}|$ may be identified with 
the physical mixing elements $|U_{\alpha 3}|~(\alpha=e,\mu,\tau).$ In view
of this, we shall consider different orderings which can give viable mixing
patterns.
\subsection{$S_\nu=Z_2$ and $T_l=Z_3$ or $Z_5$}
There are 15 different choices of $Z_2$ and 20+24 choices of the $Z_3+Z_5$ symmetry within $A_5$. Specific 
forms of $U_l$ and $U_\nu$ as discussed before can be used to obtain
$|U_{i3}|$ in all these cases. They are determined by the unknown angles
$\theta_\nu$ and $\beta_\nu$.
While the dependence of $|U_{i3}|$ on these are different for different
choices of residual symmetries all the choices share the following features
\begin{itemize}
\item  If $\beta_\nu=0$ then eq.(\ref{ulz3z5}) holds for the specific ordering
of
eigenvectors of $T_l$ as given in eq.(\ref{genmt}). The atmospheric mixing
angle is predicted to be maximal
for all the values of $\theta_\nu$. In this case, $|U_{e3}|$
is to be
identified with the 13 element of $|U|$ since $|U_{23}|=|U_{33}|$. In all
these cases, $|U_{e3}|$ depends on $\theta_\nu$ which can be chosen to
obtain
the
correct $s_{13}^2$. 
\item If $\beta_\nu\not=0$ then any of $|U_{i3}|$ can be identified with $|U_{e3}|$. It is possible in this case to choose
two unknowns $\theta_\nu$ and $\beta_\nu$ to obtain correct $\theta_{13}$
and $\theta_{23}$.
Let us discuss a specific example with  $S_\nu=E^2f_1Hf_1E$ and $T_l=E$ as
illustration. They
respectively generate $Z_2$ antisymmetry in $M_\nu$ and $Z_3$ symmetry in
$M_lM_l^\dagger$. The mixing matrix is given by $U=U_\omega^\dagger E^2f_1
V_HV_\nu$ with $V_H$ as in eq.(\ref{vh}) and $U_\omega$ as in (\ref{uw}).
The third column of the mixing matrix is then given by
\beqa \label{ui3z2}
|U_{13}|^2&=&\frac{1}{9}|c_\nu(1+2\mu_+)-s_\nu e^{i\beta_\nu} |^2~,\nonumber\\
|U_{23}|^2&=&\frac{1}{36}|-2  c_\nu (1 -\omega \mu_+)+s_\nu e^{i\beta_\nu}(\mu_++3 \omega+ \omega^2\mu_-)|^2~,\nonumber\\
|U_{33}|^2&=&\frac{1}{36}|-2  c_\nu (1 -\omega^2 \mu_+)+s_\nu e^{i\beta_\nu}(\mu_++3 \omega^2+ \omega\mu_-)|^2~.
\eeqa
For $\beta_\nu=0$, one gets $|U_{23}|^2=|U_{33}|^2$ in accordance with the
general result discussed above. In this
case, identification of $|U_{13}|^2$ with $|U_{e3}|^2$ leads to the result
$\theta_{23}=\frac{\pi}{4}$. $\theta_\nu=0.959$ then leads to $s_{13}^2\sim
0.024$. Any
of $|U_{i3}|^2$ can be identified with $|U_{e3}|^2$ when $\beta_\nu$ is
non-zero, e.g. the choice $\beta_\nu=-1.076,\theta_\nu=-0.801$ leads to
$|U_{i3}|^2=(0.444,0.024,0.532)$. In this case, $|U_{23}|^2$ plays the role
of $|U_{e3}|^2$. This specific ordering in $U$ can be obtained by exchanging 
the first and the second column of $U_\omega$.
\end{itemize}
\subsection{$S_\nu=Z_2\times Z_2$ and $T_l=Z_3$ or $Z_5$}
In this case, $S_\nu$ can be chosen in five different ways corresponding to
five different $Z_2\times Z_2$ subgroups. The corresponding neutrino mixing
matrix $U_\nu$ is given by eq.(\ref{unuz2z2}). As before $T_l$ can be chosen
in 44 different ways with $U_l$ either in ${\cal U}_3$ or ${\cal U}_5$.
Unlike in the previous case, both $U_\nu$ and $U_l$ get completely fixed
group
theoretically.  This case also predicts the maximal atmospheric mixing angle as
already outlined. Possible values of $\theta_{13}$ are also fixed.
Explicit
evaluation of various  cases reveal that in all the cases one either gets
$\theta_{13}=0$ or $s_{13}^2>0.1$. The  zero value for $\theta_{13}$
occurs for example when
$S_1=H,S_2=E^2f_2Hf_2E$ and $T_l=f_3Ef_3$. One would require relatively large
perturbations in this case to get  $\theta_{13}$  within its 3$\sigma$
range.
\subsection{$S_\nu=Z_2\times Z_2$ and $T_l=Z_2$}
This case is characterized by completely determined $U_\nu=U_\nu^{II}$ and 
$U_l=U_l^c$
containing two unknowns $\theta_l,\beta_l$. The explicit form of $U_l^c$ is
given in eq.(\ref{ulabc}) while $U_c$ can be any of the five forms given in
Table \ref{z2xz2}. $U_l$ in this case does not have the general form given
in
eq.(\ref{ulz3z5}). As a result, one does not obtain eq.(\ref{genmt})
corresponding to
the  $\mu$-$\tau$ reflection symmetry and the atmospheric 
mixing angle is not predicted to be maximal. But this case has the
following interesting feature. Explicit evaluation of
$U=U_l^{c\dagger}U_\nu^{II}$ reveals that one of the entries in the third
column of $U$ is independent of the unknown angles $\theta_l,\beta_l$ and
can be predicted group theoretically. The third column of the mixing matrix
$U$ in this case is given by
\beqa \label{upredict}
|U_{13}|^2&=& |(U_{T_l}^\dagger U_\nu^{II})_{13}|^2 ~,\nonumber\\
|U_{23}|^2&=& |c_l(U_{T_l}^\dagger
U_\nu^{II})_{23}+s_le^{-i\beta_l}(U_{T_l}^\dagger
U_\nu^{II})_{33}|^2~,\nonumber\\
|U_{33}|^2&=& |-s_l(U_{T_l}^\dagger
U_\nu^{II})_{23}+c_le^{-i\beta_l}(U_{T_l}^\dagger U_\nu^{II})_{33}|^2
\eeqa
where $T_l$ belongs to the set $O_2$ and $U_{T_l}$ to ${\cal U}_2.$
We get interesting pattern when we identify $T_l$ with $S_{1\nu}=S_1$
residing in $Z_2\times Z_2$. There exists five such choices and in all these cases,
the mixing matrix $U$ is independent of the explicit form of $U_{T_l}$. One gets from eq.({\ref{unuz2z2}) and
eq.(\ref{ulabc})
$$U=U_{23}^\dagger R_\mu  R_{12}(\frac{\pi}{4}) {\rm
diag.}(1, i,1)~.$$
The neutrino mass matrix $M_{\nu f}\equiv U_l^TM_\nu U_l$ in the flavour basis has the following form in this case:
\be \label{mnuf}
M_{\nu f}=m_0
\left(
\begin{array}{ccc}
 0 & e^{i\beta_l} c_\mu s_l-c_ls_\mu & e^{i\beta_l}c_l c_\mu
   +s_l s_\mu \\
 e^{i\beta_l} c_\mu s_l-c_ls_\mu& 0 & 0
\\
 e^{i\beta_l}c_l c_\mu
   +s_l s_\mu & 0 & 0\\
\end{array}
\right)~,\ee
where $c_l=\cos\theta_l,c_\mu=\cos\theta_\mu...$ etc.
This form can be obtained by imposing $L_e-L_\mu-L_\tau$ symmetry on $
M_{\nu f}$ as has been done in the past. 
Here, this symmetry arises as an  effective symmetry of $M_{\nu f}$ from a
very different set of basic symmetries. This symmetry leads to a degenerate
pair of neutrinos and vanishing $\theta_{13}$. The atmospheric 
mixing  angle is determined as
$\sin^2\theta_{23}=|e^{i\beta_l}c_l c_\mu
   +s_l s_\mu|^2$. Perturbations to this
symmetry have been studied in the past
\cite{Goh:2002nk,Altarelli:2005pj,Grimus:2004cj}. It is possible to
simultaneously generate the correct solar scale, solar angle and
$\theta_{13}$ with suitable but relatively large perturbations. Consider
perturbing the zero entries in eq.(\ref{mnuf}) by,
\be
\label{delmnuf}
\delta M_{\nu f}=m_0\left(
\ba{ccc}
\epsilon_1&0&0\\
0&\epsilon_2&\epsilon_4\\
0&\epsilon_4&\epsilon_3\\
\ea\right)~.\ee
Parameters $|\epsilon|$ are assumed less than the dominant entry of $M_{\nu
f}$.
We give here one example of perturbations which reproduces the observed spectrum within 3$\sigma$:

\be
\{\epsilon_1,\epsilon_2,\epsilon_3,\epsilon_4\}=\{-0.284497 ,0.284497 
,-0.0748816,0.182915\}
\ee
leading to
\be
\{\frac{\Dms}{\Dma},s_{12}^2,s_{13}^2,s_{23}^2\}=\{0.0339706,0.358739 ,
0.0243674,0.443736\}~.\ee
We have taken $\beta_l=0$ and $\cos(\theta_l-\theta_\mu)\approx -0.69$.
The values of parameters required to get above values is quite large and the
solar angle is also near to it's $3\sigma$ limit. We have verified by
randomly varying the parameters over a large range that this is a general
feature of this case. Relatively large perturbation to the basic symmetry
may come from some soft breaking as discussed for example in
\cite{Grimus:2004cj}. 

We get a non-zero $|U_{13}|^2$ when $T_l$ is not identified with $S_1$. One could determine these values
for different choices of $T_l$. 
The predicted $|U_{13}|^2$ is found from the explicit
evaluation of various  cases to take one of the three values
$~(0.095,0.25,0.65)$. Of these, only the last value provides a good leading order prediction.  $|U_{13}|^2\sim 0.65$  can be identified in this case either with $|U_{\mu
3}|^2$ or $|U_{\tau 3}|^2$
as this gives $s_{23}^2$ close to its 3$\sigma$ range $0.38-0.64$. 
\cite{Capozzi:2013csa}. This amounts to reordering of the eigenvectors of $T_l$. An example of this choice is provided by
$S_1=f_3Hf_3, S_2=E^{-1}HE$ and $T_l=f_1Hf_1$.  $|U_{i3}|^2$ are given in this case by
\beqa \label{.65}
|U_{13}|^2&=&\frac{1}{4}(2+\mu_+)\approx 0.654~,\nonumber\\
|U_{23}|^2&=&\frac{|\mu_+ c_l+2(1+\mu_+)s_l
e^{-i\beta_l}|^2}{12(2+\mu_+)}~,\nonumber\\
|U_{33}|^2&=&\frac{|-\mu_+ s_l+2(1+\mu_+)c_l
e^{-i\beta_l}|^2}{12(2+\mu_+)}~,\eeqa
One could identify  either the second or
the third entries with $s_{13}^2$
and determine $\theta_l$ accordingly, e.g. $\theta_l\sim 1.6488,\beta_l=0$
leads to $s_{13}^2\equiv |U_{33}|^2\sim 0.024$
giving $s_{23}^2c_{13}^2\equiv |U_{13}|^2\sim  0.654$. The resulting
$\sin^2\theta_{23}$ is given by $0.67$. 
Small perturbation to this case can lead to $\theta_{23}$ within 3$\sigma$
range and also split the degeneracy.
\subsection{$S_\nu=Z_2$ and $T_l=Z_2\times Z_2$}
In this case, the $Z_2$ can be generated by any of the fifteen elements in
$O_2$ while $T_l$ is generated by $T_{1l}\equiv S_1$ and $T_{2l}\equiv S_2$,
where $S_1,S_2$ form any of the five $Z_2\times Z_2$ subgroups listed in
Table \ref{z2xz2}. The PMNS matrix in this case is given by
$U=U_l^{b\dagger}
U_{S_{\nu }}V_\nu$. Just as in the previous case, the atmospheric mixing
angle
is not predicted to be maximal but now unlike it, both the angles $s_{13}^2$
and $s_{23}^2$ depend on the unknown parameters $\theta_\nu,\beta_\nu$. Not
all the choices of the residual symmetries leads to a viable values of
$\theta_{13},\theta_{23}$ in spite of the presence of the two unknowns.
We determine the allowed patterns by fitting $\theta_\nu,\beta_\nu$ to the
observed values of $\theta_{13},\theta_{23}.$
This allows us to identify cases which provide viable patterns of the mixing
angles. One finds essentially  three
patterns this way. Examples of the residual symmetries, the
patterns and best fit values of $\theta_\nu,\beta_\nu$ in each of these
cases are listed below.
\beqa
S_\nu=f_3Hf_3&: &\theta_\nu=1.42417, \beta_\nu=1.84521,s_{13}^2=0.024,
s_{23}^2=0.455,\nonumber \\
S_\nu=f_2Hf_2&: &\theta_\nu= -0.487,
\beta_\nu=0,s_{13}^2=0.0244,s_{23}^2=0.676~,\nonumber\\
S_\nu=H& :& \theta_\nu=-0.6716,\beta_\nu=
-1.1620,s_{13}^2=0,s_{23}^2=0.455~.\eeqa
All the above cases occur for the choice $T_{1l}=H$ and
$T_{2l}=E^{-1}f_2Hf_2E$. Similar results follow for different choices of
$Z_2\times Z_2$ as $T_l$ but with alternative choices of $S_\nu$. The first
case given above reproduces  the observed values of the mixing
angles $\theta_{13},\theta_{23}$. The second choice gives a $\theta_{23}$ on the verge of its
3$\sigma$ value but correct $s_{13}^2$. Thus small perturbation to this case
can lead to a viable
pattern. The third choice corresponding to $s_{13}^2=0$ would
need significant corrections from the perturbations and is analogous to the
case already discussed in section (IIIC).
\subsection{$S_\nu=Z_2\times Z_2 $ and $T_l=Z_2\times Z_2$}
In this case, the residual symmetries of neutrinos and the charged leptons correspond to (different) $Z_2\times Z_2$ groups. 
Due to the presence of two $Z_2$ groups, there are no undetermined parameters and the mixing angles $\theta_{13},\theta_{23}$
get predicted group theoretically. Since we have five different $Z_2\times Z_2$ subgroups, there are twenty different
choices which would lead to a non-trivial mixing matrix $U$. None of these
correspond to even a good zeroeth order 
values. The predicted third column of $|U|^2$ in all these cases is
\be
\label{uz2z2}
|U_{i3}|^2=
\left(
\begin{array}{c}
0.0954915\\
 0.25\\
 0.654508\\
\end{array}
\right)
~.\ee
and its permutations. These predictions are quite far from the observed mixing angles.

\section{Explicit Realization with $A_5\times Z_3$ symmetry}
We now discuss a realization of the above group theoretical discussion
choosing a specific examples of $S_\nu$ and $T_l$.  We discuss necessary
Higgs fields and  vacuum structure needed to implement above
symmetries. The model presented here is very similar to the one presented in
\cite{Joshipura:2015dsa}. The main differences being that  the neutrino
symmetry
considered in this reference is replaced by the neutrino antisymmetry.
Implementation of antisymmetry needs imposition of an additional  discrete
symmetry which we choose as $Z_3$. We use supersymmetry as basic
ingredient.

Irreducible representations (IR)  of $A_5$ are : $1+3_1+3_2+4+5$ where $3_1$
and $3_2$ are non-equivalent IR. We assign $l_L,l^c$ to $3_1$ which is explicitly generated by
 $E,H,f_1$ given in eq.(\ref{def}).
It follows from the product  rule 
$$3_1\times 3_1=(1+5)_{symm}+3_{antisym}$$
that the symmetric neutrino mass matrix can arise from $1+5$ and the charged lepton
masses can arise from all three IR. The neutrino masses are generated from 
a 5-plet $\eta_{5\nu}$ of flavon. Various fields transform under $Z_3$
as
$$(l_L,\eta_\nu)\rightarrow \omega (l_L,\eta_{5\nu})~~,~~l^c\rightarrow \omega^2 l^c~.$$
The standard Higgs  fields $H_u,H_d$ and Higgs triplet $\Delta$  are
singlets of $A_5\times Z_3$.  

The neutrino masses are generated from the following 
superpotential using the type-II seesaw mechanism:
\be \label{wnu}
W_\nu= \frac{1}{2 \Lambda} (l_L \Delta l_L)_5 h_{5\nu}
\eta_{5\nu}~.\ee
Note that the singlet term $( l_L \Delta l_L)_1$ allowed by $A_5$ is prevented above due to the $Z_3$ symmetry.
The charged lepton masses are generated by three additional flavons, a
singlet $\eta_{1l}$, a 5-plet $\eta_{5l}$ and a 3-plet $\eta_{3l}$ all
transforming trivially under $Z_3$. The
relevant superpotential is 
\be \label{wl}
W_l= \frac{1}{ \Lambda}\left[h_{sl} (l_L H_d l^c)_1 \eta_{1l} +h_{5l}(l_L
H_d
l^c)_5 \eta_{5l}+h_{3l} (l_L H_d l^c)_3
\eta_{3l}\right]~.\ee
The $Z_3$ symmetry separates the neutrino and the charged lepton sectors and does not allow flavons of one sector to
couple to the  other sector at the leading order.

We specialize to a particular choice of symmetries already discussed in
section (IIIA). This corresponds to   $T_l=E$ and $S_\nu=E^2f_1H f_1E$. 
The above  $S_\nu$ can serve as an antisymmetry of the neutrino mass matrix
if  the 5-plet $\eta_{5\nu}$ has antisymmetric vacuum expectation value:
\be \label{vacnu}
 S_\nu(5)\vev\eta_{5\nu}=-\vev\eta_{5\nu}~.\ee
$S_{\nu}(5)$ 
in eq.(\ref{vacnu}) corresponds to the five dimensional representation of
$S_\nu$. This representation can be obtained from the basic generators
defined as $a,b,c$ 
in \cite{Varzielas:2013hga} by noting the correspondence
$E=b,f_3=a$ and $H=b c$. This leads to
\be \label{snu5}
S_\nu(5)=\left(
\begin{array}{ccccc}
 \frac{1}{2} & 0 & -\frac{1}{2} & \frac{1}{2 \sqrt{2}} & \frac{\sqrt{\frac{3}{2}}}{2} \\
 0 & \frac{1}{2} & -\frac{1}{2} & -\frac{1}{\sqrt{2}} & 0 \\
 -\frac{1}{2} & -\frac{1}{2} & 0 & -\frac{1}{2 \sqrt{2}} & \frac{\sqrt{\frac{3}{2}}}{2} \\
 \frac{1}{2 \sqrt{2}} & -\frac{1}{\sqrt{2}} & -\frac{1}{2 \sqrt{2}} & -\frac{1}{4} & -\frac{\sqrt{3}}{4} \\
 \frac{\sqrt{\frac{3}{2}}}{2} & 0 & \frac{\sqrt{\frac{3}{2}}}{2} & -\frac{\sqrt{3}}{4} & \frac{1}{4} \\
\end{array}
\right) \ee
Antisymmetry of $\vev{\eta_{5\nu}}$ together with $A_5$ symmetry in $W_\nu$
results in the flavour antisymmetric mass matrix.
It is worth noting that unlike in the case of symmetry, eq.(\ref{vacnu})
breaks the symmetry $S_\nu$ completely and it does not remain as a residual
symmetry. But just as in the case with symmetry, a broken solution given in
eq.(\ref{vacnu}) may also arise from the minimization of suitable
superpotential but would need enlargement in the model. This is  explicitly demonstrated \cite{Joshipura:2015zla} in a simpler case of the group $A_4$.

Denoting the vev  $\vev{\eta_{5\nu}	}$ as $(s_1,s_2,s_3,s_4,s_5)^T$,
eq.(\ref{vacnu}) is solved for 
\be \label{solnu}
s_2=s_3-s_1~,~s_4=\sqrt{2} s_3-\frac{3 s_1}{\sqrt{2}}~,~s_5=-\sqrt{\frac{3}{2}} s_1~.\ee
Inserting this solution in eq.(\ref{wnu}), we get the neutrino mass matrix
\be
M_\nu^0=m_0\left(
\begin{array}{ccc}
\frac{-3+\sqrt{5}+y(1-\sqrt{5})}{2\sqrt{2}}&\frac{1}{\sqrt{2}
} &\frac{y}{\sqrt{2}}\\
\frac{1}{\sqrt{2}}&\frac{-2
\sqrt{5}+y(1+\sqrt{5})}{2\sqrt{2}}&\frac{y-1}{
\sqrt{2}}\\
\frac{y}{\sqrt{2}}&\frac{y-1}{\sqrt{2}}&\frac{
3+ \sqrt{5}-2 y} { 2\sqrt{2}}\\
\end{array}
\right)~.\ee
which satisfies the flavour antisymmetry, eq.(\ref{flavas}) with respect
to $S_\nu=E^2f_1Hf_1E.$ This matrix has only one complex parameter
$y\equiv \frac{s_3}{s_1}$ apart from an
overall scale.
In particular, $\tilde{M}_\nu\equiv V_{S_\nu}^T M_\nu V_{S_\nu}$ has
the form given in eq.(\ref{degtext}) with

\be \label{s3}
\tan\theta_\nu e^{i\beta_\nu}=-\frac{1+\mu_+
\frac{s_3}{s_1}}{(\mu_+-\mu_-)+\mu_-
\frac{s_3}{s_1}}~,\ee
where $V_{S_\nu}=E^2f_1V_H$ diagonalizes $S_\nu=E^2f_1Hf_1E$. The neutrino
mixing matrix is then given by $U_\nu=E^2f_1V_HV_\nu$ with $V_\nu$ as given
in eq.(\ref{unu1}) and $\theta_\nu,\beta_\nu$ given by eq.(\ref{s3}) in
terms
of
$\frac{s_3}{s_1}$. The charged lepton mixing matrix is analogously
determined by the form of $M_l$ obtained from $W_l$.
$W_l$ and the residual symmetry $T_l=E$ coincide with the one already
discussed in\cite{Joshipura:2015dsa}.
The $T_l$ invariant vacuum configuration  discussed in
 \cite{Joshipura:2015dsa} leads to the following charged lepton mass matrix 
\be \label{ml}
M_l=\left( 
\ba{ccc}
m_0&m_1-m_2&m_1+m_2\\
m_1+m_2&m_0&m_1-m_2\\
m_1-m_2&m_1+m_2&m_0\\
\ea \right) ~,\ee
where $m_{0,1,2}$ respectively label the singlet, triplet and 5-plet
contributions to $M_l$.
$M_lM_l^\dagger$ is diagonalized by the matrix (\ref{uw}) which also diagonalizes $T_l$:
$$U_\omega^\dagger M_lM_l^\dagger U_\omega={\rm diag.}(m_1^2,m_2^2,m_3^2)~$$
with eigenvalues  
\beqa
\label{evl}
\lambda_1^2&=& m_0^2+4|m_1|^2+4 m_{1R} m_0~,\nonumber \\
\lambda_2^2&=&m_0^2+|m_1|^2+3 |m_2|^2+2\sqrt{3}~Im (m_1m_2^*)-2
m_0(m_{1R}+\sqrt{3}m_{2I})~,\nonumber \\
\lambda_3^2&=&m_0^2+|m_1|^2+3 |m_2|^2-2\sqrt{3}~Im (m_1m_2^*)-2
m_0(m_{1R}-\sqrt{3}m_{2I})\eeqa
Here, $m_{1R,2R}$ and $m_{1I,2I}$ respectively denote the real and imaginary
parts of $m_{1,2}$. $m_0$ is assumed real without loss of generality.

The identification of eigenvalues $\lambda_{1,2,3}^2$ with the physical
charged lepton masses $m_{e,\mu,\tau}^2$ depends on 
the choice of parameters $m_{0,1,2}$. In particular, one can choose these
parameters in  a way that gives $\lambda_2^2=m_e^2$ ,$\lambda_{1}^2=m_\mu^2$
and $\lambda_3^2=m_\tau^2$. With this identification, 
\be \label{ul}
U_l= \frac{1}{\sqrt{3}}\left(
\ba{ccc}
1&1&1\\
\omega&1&\omega^2\\
\omega^2&1&\omega\\
\ea
\right)~.\ee
This $U_l$ together with $U_\nu=E^2f_1V_HV_\nu$ gives the mixing matrix $U$
which is already worked out in eq. (\ref{ui3z2}).
The above form of $U_l$ leads to the identification $|U_{23}|^2=s_{13}^2~,|U_{13}|^2=c_{13}^2s_{23}^2$.
Values of $\theta_\nu,\beta_\nu$ giving correct $s_{13}^2,s_{23}^2$ were
already determined in section IIIA. This 
translates to the following values of the model parameter $\frac{s_3}{s_1}$ when eq.(\ref{s3}) is used:
\be \label{s30}
\frac{s_3}{s_1}\approx 0.9979 e^{-0.7181 i}~.\ee
Non-zero neutrino masses remain degenerate at the leading order. They can be split and the solar angle can be determined
by perturbations which break antisymmetry at the non-leading order. A simple perturbation can be generated by introducing a singlet flavon 
$\eta_{1\nu}$ transforming as $\eta_{1\nu}\rightarrow \omega^2\eta_{1\nu}$
under $Z_3$. This flavon leads to a non-leading term 
$$ \frac{h_{1\nu}}{2 \Lambda^2} (l_l \Delta l_L)_1 
\eta_{1\nu}^2 $$
in $W_\nu$. This generates a diagonal perturbation which can be
parameterized as
$$M_\nu=m_0(\hat{M}^0_\nu+\epsilon I)~$$
with 
$\hat{M}^0_\nu\equiv\frac{M^0_\nu}{m_0}$ and $|\epsilon|<<1$.
This simple perturbation is enough to generate the
solar splitting without disturbing the zeroeth order
values of $s_{13}^2,s_{23}^2$ significantly. One could vary $\frac{s_3}{s_1}$ around
the zeroeth order value determined in eq.(\ref{s30}) and find the region of parameters which fits
the data with $|\epsilon|<0.1$. This procedure leads to a solution close to
the best fit values of all parameters. For example,
$\frac{s_3}{s_1}=1.00019 e^{-0.711498 i}$ and $\epsilon=0.0168241$ leads to
\beqa \label{results}
\sin^2\theta_{12}=0.295455&~,~&\sin^2\theta_{13}=0.0235172~,\nonumber \\
~\sin^2\theta_{23}=0.449634&~,~&\frac{\Dms}{\Dma}= 0.0285398 ~.\eeqa
\section{Summary}
We have studied consequences of an ansatz of flavour antisymmetry in the context of the flavour group $A_5$
assuming that $S_\nu$ in eq.(\ref{flavas}) and $T_l$ in eq.(\ref{mlsym}) are
contained in the group $A_5$. These assumptions constraint the mixing
pattern which we have determined in various cases. The use of flavour
antisymmetry in the context
of the $A_5$ group necessarily leads to a degenerate pair of neutrinos in
addition to a 
massless one. This is a good zeroeth order prediction. Small perturbations
splitting the degeneracy can lead to a viable neutrino masses. The 
predicted neutrino mass hierarchy is inverted.

We have considered discrete subgroups  $Z_2$ and $Z_2\times Z_2$ of $A_5$ as residual symmetries of $M_\nu$
and discrete groups $Z_3,Z_5,Z_2$ and $Z_2\times Z_2$ contained in $A_5$ as symmetries of $M_lM_l^\dagger$
and worked out the resulting mixing patterns at the leading order in all the
cases. The third column of the mixing matrix and hence the angles
$\theta_{13},\theta_{23}$ get determined at this order. Various predictions
discussed in section III can be summarized as follows:
\begin{itemize}
\item It is possible to get a universal prediction of the maximal
atmospheric mixing angle with the choice 
$S_\nu$ as $Z_2$ or $Z_2\times Z_2$  and $T_l$ as any element in $Z_3,Z_5$.
For $S_\nu=Z_2$, one can also get the correct $\theta_{13}$ at the leading
order while the case of $S_\nu=Z_2\times Z_2$ predicts either
$\theta_{13}=0$ or large $s_{13}^2\geq 0.1$. 
\item The case $T_l=Z_2$ and $S_\nu=Z_2\times Z_2$ does not predict maximal
$\theta_{23}$ but can be used to predict one of the entries of the third
column. The other entry gets determined by an unknown angle inherent with
the use of the  $Z_2$ groups. The viable predictions within this case are either
$\theta_{13}=0$ or $s_{23}c_{13}^2=0.65$. The former requires large
perturbation and we have presented a typical set of such perturbation which
lead to correct description of masses and mixing angle.
\item The case $S_\nu=Z_2$ and $T_l=Z_2\times Z_2$ involves an unknown angle
and a phase. Not all possible choices of $S_\nu,T_l$ in this category  can
lead to correct mixing  in spite of the presence of two unknowns. We have
identified cases which lead to correct description of the mixing angles
$\theta_{13},\theta_{23}$.
\item The case of both $S_\nu$ and $T_l$ belonging to different $Z_2\times
Z_2$ subgroups of $A_5$ is fully predictive without any unknowns. But none
of the possible cases within this category lead even to a good zeroeth order
prediction.
\end{itemize}

We have supplemented the group theoretical derivation of  
the mixing patterns in $A_5$ with a concrete example. We have determined the
Higgs content and the required vacuum pattern which realizes one of the
viable cases discussed group theoretically. The use of a concrete model also
allows a systematic discussion of possible perturbations and we have given an
example of a perturbation within the model which can be used to
split the degeneracy of neutrinos and which can give the correct descriptions of all mixing angles and masses.
\section{Acknowledgement}
Work of ASJ was supported 
by BRNS (DAE) through the Raja Ramanna Fellowship. 
  
\bibliography{flavasa5}
\bibliographystyle{apsrev4-1}
\end{document}